# Analyses of Flight Time During Solar Proton Events and Solar Flares

X. H. Xu[1], Y. Wang[1], F. S. Wei[1], X. S. Feng[1], M. H. Bo[2], H.W. Tang[2], D. S. Wang[2], B. Lei[2], B.Y. Wang[1], P. B. Zuo[1], C. W. Jiang[1], X.J. Xu[3], Z. L. Zhou[3], Z. Li[4], P. Zou[1], L. D. Wang[1], Y. X. Gu[1], Y. L. Chen[1], W. Y. Zhang[1], P. Sun[1]

1 Institute of Space Science and Applied Technology, Harbin Institute of Technology, Shenzhen, 518055, China
2 Travelsky Mobile Technology Limited, Beijing, 100087, China
3 State Key Laboratory of Lunar and Planetary Sciences, Macau University of Science and Technology, Macao, China
4 Nanjing University of Information Science and Technology, Nanjing, China

**Abstract**

Analyzing the effects of space weather on aviation is a new and developing topic. It has been commonly accepted that the flight time of the polar flights may increase during solar proton events because the flights have to change their route to avoid the high-energy particles. However, apart from such phenomenon, researches related to the flight time during space weather events is very rare. Based on the analyses of 39 representative international air routes around westerlies, it is found that 97.44% (94.87%) of the commercial airplanes on the westbound (eastbound) air routes reveal shorter (longer) flight time during solar proton events compared to those during quiet periods, and the averaged magnitude of change in flight time is ~10 min or 0.21%-4.17% of the total flight durations. Comparative investigations reassure the certainty of such phenomenon that the directional differences in flight time are still incontrovertible regardless of over-land routes (China-Europe) or over-sea routes (China-Western America). Further analyses suggest that the solar proton events associated atmospheric heating will change the flight durations by weakening certain atmospheric circulations, such as the polar jet stream. While the polar jet stream will not be obviously altered during solar flares so that the directional differences in flight time are not found. Besides the conventional space weather effects already known, this paper is the first report that indicates a distinct new scenario of how the solar proton events affect flight time. These analyses are also important for aviation since our discoveries could help the airways optimize the air routes to save passenger time costs, reduce fuel costs and even contribute to the global warming issues.

# Introduction

Space weather refers to conditions on the sun and in the solar wind, magnetosphere,

and ionosphere that can influence the performance and reliability of space-borne and ground-based technological systems[1,2]. Space weather also has very important impacts on human society including communications, navigation, and electric power grids[3]. In recent years, the relationships between space weather and aviation become new and developing research hotspots, and researching the impacts of space weather on the aviation industry has attracted more and more scholars and international communities[4-7]. Solar proton events (SPEs) and solar flares are two typical space weather events. SPEs are very high-energy protons coming from the sun which are commonly believed to have greater influences on the higher latitudes, since the high-energy protons will be deflected by the geomagnetic field[8,9]. While solar flares behave as increased electromagnetic radiations directly propagating to the Earth, and their influences are generally regarded to concentrated at the lower latitudes[10,11].

The impacts of space SPEs on flight times have been mentioned in some literature[12,13]. It is widely known that SPEs associated ionizing radiation and single-event error at conventional cruising altitude could be harmful to persons and electronics [14,15]. Hence, for safety concerns, some polar flights have to change their original routes to avoid such effects and the flight time will increase [16,17]. Solar flares have also been demonstrated to be able to cause flight delays[6,7,18] but no one knows whether they can affect flight time.

Actually, the impacts of SPEs on Earth are not confined to the above referred effects. Many existing studies have suggested that SPEs can result in warming at high latitudes[19-21], which will weaken the meridional temperature gradient and change the meridional pressure gradient, and accordingly, the zonal wind will slow down and the polar jet stream could be weakened [22-24]. Meanwhile, the polar jet stream is very important for aviation since it can affect airplanes at cruise altitudes to increase (headwind) or decrease (downwind) the flight time[25-27]. Such effects are aviation general knowledge and many eastbound flights usually have shorter flight time than the westbound flights on the same route. Therefore, we believe that the underlying mechanism of how SPEs influence the flight time could beyond the conventional understandings[12,13]. To reveal such a new scenario, 39 representative international air routes around westerlies are selected to investigate the flight time of commercial airplanes during SPEs and solar flares. It is found that most of the airplanes reveal systematically changes in flight time during SPEs compared to those during quiet periods, while such phenomenon is not seen during solar flares.

**Data and Methods**

The flight data used in the analyses is provided by Travelsky Mobile Technology Limited, an affiliated company of Civil Aviation Administration of China. The intact flight data is only available from 2015, and to avoid the impacts of covid-19 on aviation, the data after 2020 is not used. Finally, detailed flight data from January 1st, 2015 to December 31th, 2019 is used in the statistics. Since the flight time is thought to be related to the polar jet stream, to reveal the impacts of SPEs and solar flares on flight times more clearly, the selection of air routes has to meet some requirements. First of all, the air routes should near the polar jet stream so they can be affected by it. In addition, the air routes should not be too short or the influences of polar jet stream will be negligible. Last but not least, the selected air routes should contain enough flight records to ensure the credibility of statistical results. Finally, as seen in fig.1, 39 representative international air routes are selected and each of them contains more than 1,000 flight records.

SPE list are obtained from the National Oceanic and Atmospheric Administration (NOAA) Space Weather Prediction Center[28]. Since the SPEs usually last 2-3 days[29], the SPE-affected periods are defined from the beginning of the event to the next 72 hours. The solar flare events are also mainly obtained from NOAA website [30], and we also use the soft X-ray data from GOES satellites to complement the unlisted events. Only M-class and X-class solar flare events that are not disturbed by other space weather events (e.g., Coronal Mass Ejections, SPEs) are used. The durations of all solar flare events are set to 24 hours to avoid the interference of the 24-hour intrinsic period of flight operation[6,7]. Meanwhile, we define the quiet periods as the entire days (from 00:00 to 24:00) when there are no space weather events[6,7]. Finally, 52 solar flare events and 8 SPEs are selected from January 1, 2015 to December 30, 2019, and the total number of the related individual flight records is 8,323,984.

The polar jet stream is not fixed on Earth but varies with seasons, climate and other factors[31], and the flights on the 39 selected air routes will not be affected all the time during their voyages. Hence, to have more intensive and comprehensive understandings to the flight time on the selected air routes, we would like to classify the 39 routes into two groups: 27 over-land routes (group 1: China-Europe) and 12 over-sea routes (group 2: China-Western America). Such comparative classification could help eliminate other confounding factors as much as possible. For all of the 39 selected air routes, the 'Westbound' used here means all the flights flying from Western America to China and from China to Europe whose voyages will mainly against the polar jet stream, while 'Eastbound' on the contrary.

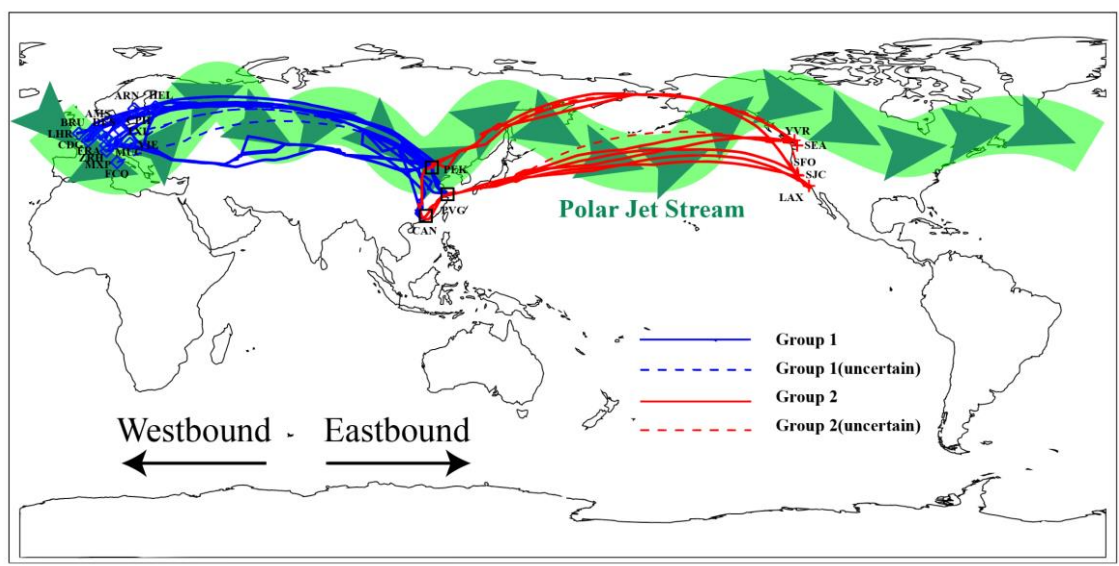

Fig. 1 The typical trajectories of the 39 selected air routes and the approximate location of the polar jet stream. The information of the air routes are obtained from FlightAware[32], while the uncertain routes (or no route data) are shown in dashed lines.

## Results

The polar jet stream around westerlies plays an important role in commercial aircraft operations because it could create strong headwinds or tailwinds over busy mid-latitude air routes and influence the flight time of commercial flights[26]. While it is inferred that polar jet stream will be weakened during SPEs[21-23,33] so that the flight time during SPEs will be altered. First and foremost, the flight time of the 39 air routes during SPEs and quiet periods are analyzed. As seen in fig.2, compared to the quiet periods, flights on 38 out of 39 westbound air routes, flying mainly against the polar jet stream, show decreased flight time during SPEs. While on the contrary, flights on 37 out of 39 eastbound air routes reveal increased flight time. To reassure the certainty of such phenomenon, 27 over-land routes and 12 over-sea routes are investigated respectively, and the results are still the same. The directional differences in flight time are so obvious and ubiquitous that such phenomena suggest that there should be overall intrinsic relationships exist between the flights and SPEs but not just statistical anomaly.

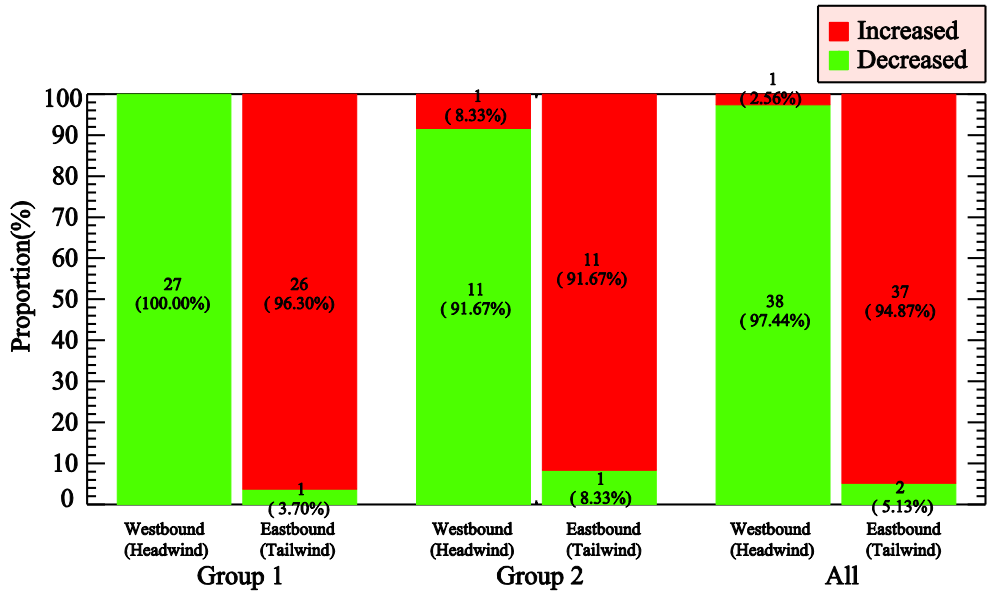

Fig. 2 Flight time changes of over-land routes (left), over-sea routes (middle) and all routes (right) during SPEs compared to those during quiet periods. The numbers of the routes and their percentage are marked on the percent bar graph. Red (green) color represents that the flight time during SPEs is longer (shorter) than those during quiet periods. Westbound (Eastbound) represents the flights flying mainly against (follow) the polar jet stream.

Although the overall results of the directional differences in flight time shown in Fig.2 seem incontrovertible, it is still wondered whether such effect is substantial or caused by some outliers on a certain air route. It is revealed in Fig.3 (a) that the distribution of flight time on a westbound route (PVG to CDG) during SPEs systematically shifts to the left as a whole compared to those during quiet periods. This result means that the flight time during SPEs will generally decrease, and the averaged flight time difference is ~10.69 min. Such systematical changes in flight time can also be verified in Fig.3 (b) which displays an eastbound route (MUC to PEK) with a ~13.29 min averaged time difference. Moreover, Fig.3 (c) and Fig.3 (d) provide the histograms of flight time difference between those during SPEs and quiet periods for all westbound and eastbound flights. One can see that the flight time difference ranges from -25 min to 10 min for westbound flights (-5 min to 30 min for eastbound flights), or 0.21%-4.17% of the total flight durations. These results not only consist of the previous statistics but also prove the systematical modulating effects of SPEs. Moreover, they could imply that the decreased (increased) flight time during SPEs would be closely related to the decreased headwind (tailwind) effect resulting from the weakening of the polar jet stream.

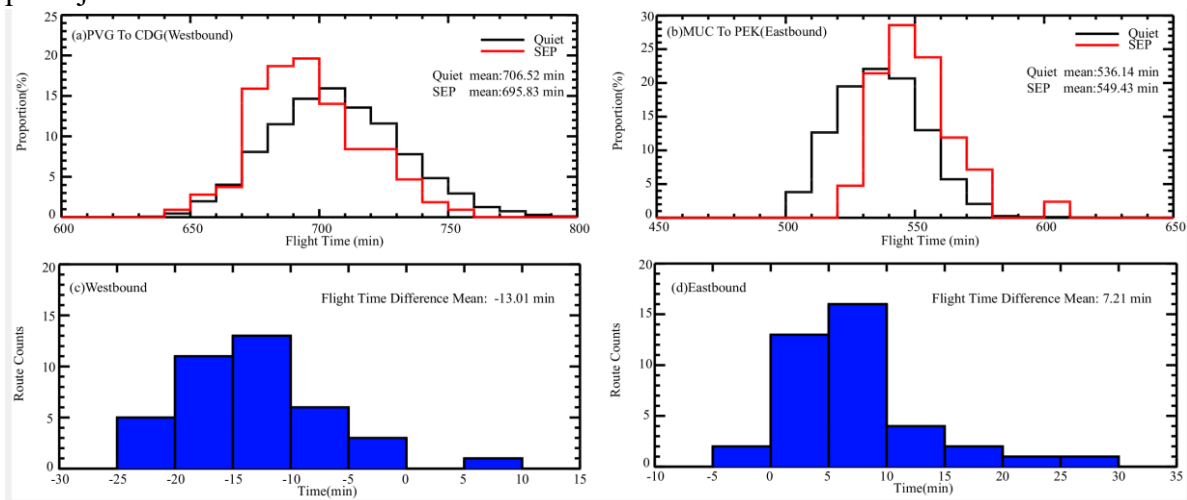

Fig. 3 (a) Distribution of flight time on a westbound route (PVG to CDG) during SPEs (red) and quiet periods (black); (b) Distribution of flight time on an eastbound route (MUC to PEK) during SPEs and quiet periods; (c) Histogram of flight time difference between those during SPEs and quiet periods for all westbound flights. (d) Histogram of flight time difference between those during SPEs and quiet periods for all eastbound flights.

## Discussions and Conclusions

As stated in the introduction, it has long been known that the SPEs associated ionizing radiations or single-event error will force some polar flights to change their routes so that the flight time will increase. However, in addition to such safety concerns, our presented results reveal a brand new and very interesting scenario of how SPEs can affect flight time. Actually, the relevant investigations on the warming effects of SPEs and how the meridional temperature gradients would affect polar jet stream have been revealed for years [21,33-35], while the direct relationship between SPEs and polar jet stream has not been revealed. Most importantly, no one has realized the possibility that SPEs can influence flight time by altering the atmospheric circulations.

It is inferred that when SPEs reach Earth, the geomagnetic field will deflect some high-energy particles to high latitudes and particle precipitation will occur[36]. Many simulations[33,37] and observations [38,39] both suggest that particle precipitation will heat the atmosphere and warm the Eurasia and North American. In particular, some extreme SPEs, as those during Carrington Event, could even lead the surface temperature warming up to 7K in Europe and Russia[21]. While the difference in temperature between the equator region and polar region is crucial for the polar jet stream. The meridional temperature gradients and pressure gradients will be weakened due to the temperature rise at high latitudes, so that the zonal wind will be slowed down and the polar jet stream will be weakened in mid-latitude[31,34]. Such phenomenon can be observed not only in winter when the equatorial to polar meridional temperature gradient is large, but also in summer when the temperature gradient is relatively small[40]. In such cases, the airplanes flying westbound (eastbound) will encounter weaker (stronger) air resistance and the flight time would probably decrease (increase).

Since the space weather events associated magnetospheric-ionospheric disturbances can also have remarkable influences on aviation[6,7]. It is still necessary to confirm if the changes in flight time are caused by the conventional space weather effects. Therefore, similar statistics are carried out by using the solar flare events instead. While the significant directional differences in flight time during solar flares should not be expected, because the solar flares are electromagnetic radiations which will not heat the polar region to weaken the meridional temperature gradients so that the weakened polar jet stream will not occur. The final results can be seen in Fig. 4, it is shown that more than half of the flights will increase their flight time regardless of westbound or eastbound. However, the directional differences in flight time during solar flares is completely absent. In another way, such results also imply that the SPEs caused directional differences in flight time are not resulted from the conventional space weather effects.

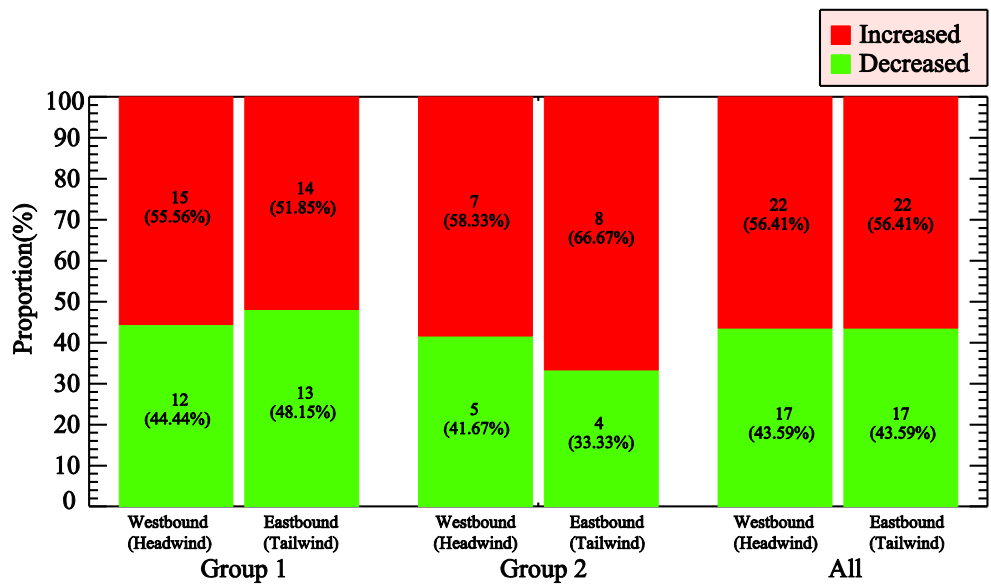

Fig. 4 Flight time changes of over-land routes(left), over-sea routes(middle) and all routes (right) during solar flares compared to those during quiet periods. The numbers of the routes and their percentage are marked on the percent bar graph. Red (green) color represents that the flight time during solar flares is longer (shorter) than those during quiet periods. Westbound (Eastbound) represents the flights flying mainly against (follow) the polar jet stream.

To summarize, the effects of space weather on Earth could go far beyond human understandings. By analyzing 39 representative international air routes around westerlies, an interesting phenomenon about the directional differences in flight time is first revealed during SPEs. Compared to those during quiet periods, our investigations show that most of the flights on the westbound (eastbound) air routes reveal shorter (longer) flight time during SPEs, and the averaged magnitude of change in flight time is ~10 min or 0.21%-4.17% of the total flight durations. Such directional difference is obvious and ubiquitous regardless of over-land routes or over-sea routes. The intrinsic relationships between the flight time and SPEs is discussed that the SPEs associated polar heating will weaken the meridional temperature gradients to slow down the polar jet stream so that alter the flight time. While comparative analyses of solar flare related flights are also carried out and the directional differences are not shown. These results expand our understandings to conventional space weather effects, and the related analyses are also important for aviation since these discoveries could help optimize the air routes to save passenger time costs, reduce fuel costs and even contribute to the global warming issues.

## Acknowledgements

This work is jointly supported by the National Natural Science Foundation of China (41731067 and 42174199), Guangdong Basic and Applied Basic Research Foundation (2021A1515012581), and the Shenzhen Technology Project (GXWD20201230155427003-20200804210238001).

# References

1. Schwenn, R. Space Weather: The Solar Perspective. *Living Reviews in Solar Physics* **3**, 2, doi:10.12942/lrsp-2006-2 (2006).
2. Pulkkinen, T. Space Weather: Terrestrial Perspective. *Living Reviews in Solar Physics* **4** (2007).
3. Hapgood, M. Prepare for the coming space weather storm. *Nature* **484**, 311-313, doi:10.1038/484311a (2012).
4. Hubert, G. & Aubry, S. Study of the Impact of Past Extreme Solar Events on the Modern Air Traffic. *Space Weather* **19**, e2020SW002665, doi:https://doi.org/10.1029/2020SW002665 (2021).
5. Jones, J. B. L. *et al.* Space weather and commercial airlines. *Advances in Space Research* **36**, 2258-2267, doi:10.1016/j.asr.2004.04.017 (2005).
6. Xu, X. H. *et al.* The Effects of Space Weather on Flight Delays. *Submitted* (2022).
7. Wang, Y. *et al.* Characteristics of Flight Delays during Solar Flares. *Submitted* (2022).
8. Bailey, D. K. Polar-CAP Absorption. *Planetary and Space Science* **12**, 495-541, doi:10.1016/0032-0633(64)90040-6 (1964).
9. Jackman, C. H. *et al.* Northern Hemisphere atmospheric effects due to the July 2000 solar proton event. *Geophysical Research Letters* **28**, 2883-2886, doi:10.1029/2001gl013221 (2001).
10. Qian, L. Y., Burns, A. G., Chamberlin, P. C. & Solomon, S. C. Variability of thermosphere and ionosphere responses to solar flares. *J Geophys Res* **116**, A10309, doi:10.1029/2011ja016777 (2011).
11. Pettit, J. *et al.* Effects of the september 2005 Solar Flares and Solar Proton Events on the Middle Atmosphere in WACCM. *J Geophys Res* **123**, 5747-5763, doi:10.1029/2018ja025294 (2018).
12. Saito, S., Wickramasinghe, N. K., Sato, T. & Shiota, D. Estimate of economic impact of atmospheric radiation storm associated with solar energetic particle events on aircraft operations. *Earth Planets and Space* **73**, 1-10 (2021).
13. Fujita, M., Sato, T., Saito, S. & Yamashiki, Y. Probabilistic risk assessment of solar particle events considering the cost of countermeasures to reduce the aviation radiation dose. *Scientific Reports* **11**, 17091, doi:10.1038/s41598-021-95235-9 (2021).
14. Lanzerotti & Louis, J. Space Weather: Historical and Contemporary Perspectives. *Space Science Reviews* **212**, 1253-1270, doi:10.1007/s11214-017-0408-y (2017).
15. Knipp, D. J. Essential science for understanding risks from radiation for airline passengers and crews. *Space Weather-the International Journal of Research & Applications* **15**, 549-552, doi:10.1002/2017SW001639 (2017).
16. Meier, M. M. & Matthi, D. A space weather index for the radiation field at aviation altitudes. *Journal of Space Weather & Space Climate* **4**, A13, doi:10.1051/swsc/2014010 (2014).
17. Matthi, D., Schaefer, M. & Meier, M. M. Economic impact and effectiveness of radiation protection measures in aviation during a ground level enhancement. *J. Space Weather Space Clim.* **5(2015)**, 1227-1234, doi:10.1051/swsc/2015014 (2015).
18. Marque, C. *et al.* Solar radio emission as a disturbance of aeronautical radionavigation. *Journal of Space Weather and Space Climate* **8**, 1-13, doi:10.1051/swsc/2018029 (2018).
19. Baumgaertner, A. J. G., Seppala, A., Jockel, P. & Clilverd, M. A. Geomagnetic activity related NOx enhancements and polar surface air temperature variability in a chemistry climate model: modulation of the NAM index. *Atmos. Chem. Phys.* **11**, 4521-4531, doi:10.5194/acp-11-4521-2011 (2011).
20. Sukhodolov, T. *et al.* Atmospheric impacts of the strongest known solar particle storm of 775 AD. *Scientific Reports* **7**, 45257, doi:10.1038/srep45257 (2017).

21. Calisto, M., Usoskin, I. & Rozanov, E. Influence of a Carrington-like event on the atmospheric chemistry, temperature and dynamics: revised. *Environmental Research Letters* **8**, 045010, doi:10.1088/1748-9326/8/4/045010 (2013).
22. Francis, J. A. & Vavrus, S. J. Evidence linking Arctic amplification to extreme weather in mid-latitudes. *Geophysical Research Letters* **39**, L06801, doi:10.1029/2012gl051000 (2012).
23. Cohen, J. *et al.* Recent Arctic amplification and extreme mid-latitude weather. *Nature Geoscience* **7**, 627-637, doi:10.1038/ngeo2234 (2014).
24. Lee, S. H., Williams, P. D. & Frame, T. H. A. Increased shear in the North Atlantic upper-level jet stream over the past four decades. *Nature* **572**, 639-642, doi:10.1038/s41586-019-1465-z (2019).
25. Arrow, K. J. On the Use of Winds in Flight Planning. *Journal of Meteorology* **6**, 150-159, doi:10.1175/1520-0469(1949).
26. Williams, P. D. Transatlantic flight times and climate change. *Environmental Research Letters* **11**, 024008, doi:10.1088/1748-9326/11/2/024008 (2016).
27. Karnauskas, K. B., Donnelly, J. P., Barkley, H. C. & Martin, J. E. Coupling between air travel and climate. *Nature Climate Change* **5**, 1068-1073, doi:10.1038/nclimate2715 (2015).
28. NOAA-SPEs. *https://umbra.nascom.nasa.gov/SEP/*
29. Reames, D. V. Particle acceleration at the Sun and in the heliosphere. *Space Science Reviews* **90**, 413-491, doi:10.1023/a:1005105831781 (1999).
30. NOAA-Flares. *https://www.ngdc.noaa.gov/stp/space-weather/solar-data/solar-features/solar-flares/x-rays/goes/xrs/*
31. Hall, R., Erdelyi, R., Hanna, E., Jones, J. M. & Scaife, A. A. Drivers of North Atlantic Polar Front jet stream variability. *International Journal of Climatology* **35**, 1697-1720, doi:10.1002/joc.4121 (2015).
32. FlightAware. *https://flightaware.com/*, <https://flightaware.com/> (
33. Rozanov, E., Calisto, M., Egorova, T., Peter, T. & Schmutz, W. Influence of the Precipitating Energetic Particles on Atmospheric Chemistry and Climate. *Surveys in Geophysics* **33**, 483-501, doi:10.1007/s10712-012-9192-0 (2012).
34. Francis, J. A. & Vavrus, S. J. Evidence for a wavier jet stream in response to rapid Arctic warming. *Environmental Research Letters* **10**, 014005, doi:10.1088/1748-9326/10/1/014005 (2015).
35. Screen, J. A. CLIMATE SCIENCE Far-flung effects of Arctic warming. *Nature Geoscience* **10**, 253-254, doi:10.1038/ngeo2924 (2017).
36. Kress, B. T., Mertens, C. J. & Wiltberger, M. Solar energetic particle cutoff variations during the 29-31 October 2003 geomagnetic storm. *Space Weather* **8**, 1-13, doi:10.1029/2009sw000488 (2010).
37. Rozanov, E. *et al.* Atmospheric response to NOy source due to energetic electron precipitation. *Geophysical Research Letters* **32**, L14811, doi:10.1029/2005gl023041 (2005).
38. Seppala, A., Randall, C. E., Clilverd, M. A., Rozanov, E. & Rodger, C. J. Geomagnetic activity and polar surface air temperature variability. *J Geophys Res* **114**, A10312, doi:10.1029/2008ja014029 (2009).
39. Maliniemi, V., Asikainen, T., Mursula, K. & Seppala, A. QBO-dependent relation between electron precipitation and wintertime surface temperature. *J. Geophys. Res.-Atmos.* **118**, 6302-6310, doi:10.1002/jgrd.50518 (2013).
40. Coumou, D., Di Capua, G., Vavrus, S., Wang, L. & Wang, S. The influence of Arctic amplification on mid-latitude summer circulation. *Nature Communications* **9**, 2959, doi:10.1038/s41467-018-05256-8 (2018).